\documentclass[twocolumn,3p,times,procedia]{elsarticle}
\usepackage{ecrc}
\usepackage{graphicx}
\usepackage{titlesec}
\usepackage[utf8]{inputenc}
\usepackage{subcaption}
\usepackage{textcomp}
\usepackage{amssymb}
\usepackage{amsmath}
\usepackage{multicol}
\usepackage{epsfig,amssymb}
\usepackage{epstopdf}
\usepackage{lipsum}
\usepackage{float}
\usepackage{array}
\usepackage{stackengine}

\usepackage{url} 
\usepackage{color}
\usepackage{threeparttable}
\volume{00}

\newcolumntype{x}[1]{>{\vspace{0pt}\hspace{0pt}}p{#1}}%

\firstpage{1}

\runauth{}

\jid{Digit.Commun.Netw.}

\usepackage[figuresright]{rotating}
\usepackage{bm}
\usepackage{caption}

\usepackage{fancyhdr}

\fancyheadoffset[RO,EL]{0pt}
\usepackage{geometry}

\begin{document}

\begin{frontmatter}

\dochead{}

\title{
\begin{flushleft}
{\bf \Huge Blockchain-enabled Resource Management and Sharing for 6G Communications}
\end{flushleft}
}


\author[]{\bf \Large \leftline {Hao Xu $^a$, Paulo Valente Klaine $^a$, Oluwakayode Onireti $^a$, Bin Cao $^b$,} \bf \Large \leftline {Muhammad Imran $^a$, Lei Zhang $^*$$^a$}}

\address{\bf  \leftline {$^a$James Watt School of Engineering, University of Glasgow, Scotland, United Kingdom}
}
\address{\bf  \leftline {$^b$Beijing University of Posts and Telecommunications of China, Beijing, China}
}

\fntext[]{Hao Xu, Paulo Valente Klaine, Oluwakayode Onireti, Muhammad Imran and Lei Zhang are with James Watt School of Engineering, University of Glasgow (email: h.xu.2@research.gla.ac.uk; [Paulo.ValenteKlaine, Oluwakayode.Onireti, Muhammad.Imran, Lei.Zhang]@glasgow.ac.uk).  This work was supported in partby the U.K. EPSRC (EP/S02476X/1)}

\fntext[]{Bin Cao is with Beijing University of Posts and Telecommunications of China (email: caobin65@163.com ).}
\cortext[]{Corresponding author: Lei Zhang}
\begin{abstract}
The sixth generation (6G) network must provide performance superior to previous generations in order to meet the requirements of emerging services and applications, such as multi-gigabit transmission rate, even higher reliability, sub 1 millisecond latency and ubiquitous connection for Internet of Everything.  However, with the scarcity of spectrum resources, efficient resource management and sharing is crucial to achieve all these ambitious requirements. One possible technology to enable all of this is blockchain, which has recently gained significance and will be of paramount importance to 6G networks and beyond due to its inherent properties. In particular, the integration of blockchain in 6G will enable the network to monitor and manage resource utilization and sharing efficiently. Hence, in this article, we discuss the potentials of blockchain for resource management and sharing in 6G using multiple application scenarios namely, Internet of things, device-to-device communications, network slicing, and inter-domain blockchain ecosystems. 

\end{abstract}


\end{frontmatter}


\section{Introduction}
\label{sec:Intro}
As of last year, the fifth generation of mobile networks, 5G, is already being commercialized in some parts of the world, with the expectation of addressing limitations of current cellular systems as well as providing an underlying platform for new services to emerge and thrive~\cite{Andrews2014}.
5G was envisioned to be not only a faster 4G, but also an enabler for several other applications, such as the Internet of Everything (IoE), industry automation, intelligent transportation and remote healthcare, to name a few, by providing ultra high reliability, latency as low as 1 millisecond, and increased network capacity and data rates~\cite{Saad2019}.
However, despite the emergence of new technologies, such as millimeter waves, massive Multiple-Input-Multiple-Output (MIMO) and the utilization of higher frequency bands, it is clear that 5G will not be able to attend all of these requirements, albeit improving significantly from its predecessors.
As such, research has already shifted towards the next generation of mobile networks, 6G~\cite{Saad2019,Tariq2019,Zhang2019,Yang2019}.

It is expected that by 2030 our society will shift towards a more digitized, data driven and intelligently inspired society, that needs a near-instant and ubiquitous wireless connectivity~\cite{Zhang2019,David2018}.
Thus, several novel applications that provide such interaction and integration are bound to emerge in the next decade~\cite{Zhang2019}.
As such, some key trends that are foreseen to emerge in the near future are: virtual and augmented reality, 8K video streaming, holograms, remote surgery, the industry 4.0, smart homes, fog computing, artificial intelligence integrated services, unmanned aerial vehicles, and autonomous vehicles, to name a few~\cite{Zhang2019,Yang2019,Chen2020}.
These, by their turn, will demand much more from mobile networks in terms of reliability, latency and data rates than 5G and its improvements can support~\cite{Saad2019,Zhang2019,Yang2019}.
As such, several research initiatives around the globe are already working towards shaping the direction of 6G, and some of its key requirements are already being speculated, as~\cite{Saad2019,Tariq2019,Zhang2019}:
	\begin{itemize}
		\item Provide peak data rates of at least $1$ Tb/s and latency of less than 1ms;
		\item Support user mobility up to 1000 km/h;
		\item Operate in GHz to THz frequency range;
		\item Increase the network spectral efficiency, energy efficiency and security;
		\item Harness the power of big data, enabling a self-sustaining wireless network;
		\item Support for a massive number of devices and things, enabling the IoE.
	\end{itemize}

To enable all of the above and increase the systems total capacity, two different approaches are possible, according to Shannon's information theory: either increase the system bandwidth or improve the spectral efficiency~\cite{Zhang2019,Kotobi2018,Weiss2019}.
It is well-known that spectrum management is a key to achieve efficient spectrum usage and it still has issues.
For example, it is known that current fixed paradigms for spectrum assignment and resource management is a major challenge in mobile networks.
This will become even more challenging in 6G, due to the ever-growing number of subscribers and their need of intermittent connectivity, as well as the development of more data-hungry applications.
Moreover, a number of studies have shown that fixed spectrum allocation, despite being less complex, produces low spectrum efficiency, since license holders of that spectrum do not utilize it all the time (see~\cite{Kotobi2018} and references therein).

As such, several approaches have been proposed to improve spectrum management, such as Opportunistic Spectrum Access (OSA) or auction mechanisms.
Despite the benefits of these approaches, issues in terms of security, high computational power and convergence, are present.
On top of that, even if such protocols provide some collaborations at the system level, collaboration between users is still not considered, hindering the overall performance of those solutions.
Since 6G is expected to be much more cooperative than its preceding generations, with new technologies such as wireless power transfer, mobile edge computing, the IoE and Device-to-Device (D2D) communications, heavily relying on the cooperation between devices, novel approaches that do not rely on a central authority controlling spectrum and resource management, such as blockchain, are needed~\cite{Tariq2019,Saad2019}.
	
Due to its inherit characteristics, blockchain is being regarded as the next revolution in wireless communications, with even the Federal Communications Commission (FCC) emphasizing the crucial role that it can play in 6G and beyond~\cite{Dai2019}.
The main idea behind blockchain is that of an open and distributed database (ledger), where no single party has control and transactions\footnote{These transactions can mean anything, such as holdings of a digital currency (i.e. Bitcoin), movement of goods across a supply chain, spectrum and resource allocation in wireless networks, etc.~\cite{Weiss2019}.} are securely recorded in blocks.
Each block is chained together to its predecessor in a sequential, verified and secure manner, without the need of a trusted third party.
As such, blockchain is expected to revolutionize resource and spectrum sharing by eliminating the central authority and replacing it with a distributed one, enabling the trade of assets without them reaching the central server, improving network security and reducing its cost~\cite{Dai2019Survey, Sun2019b}.
	
This integration between wireless networks and blockchain will allow the network to monitor and manage spectrum and resource utilization in a more efficient manner, reducing its administration costs and improving the speed of spectrum auction.
In addition, due to its inherit transparency, blockchain can also record real-time spectrum utilization and massively improve spectrum efficiency by dynamically allocating spectrum bands according to the dynamic demands of devices~\cite{Weiss2019}.
Moreover, it can also provide the necessary but optional incentive for spectrum and resource sharing between devices, fully enabling new technologies and services that are bound to emerge~\cite{Sun2019b}.
Furthermore, with future wireless networks shifting towards decentralized solutions, with thousands of cells deployed by operators and billions of devices communicating with each other, fixed spectrum allocation and operator controlled resource sharing algorithms will not be scalable nor effective in future networks.
As such, by designing a communications network coupled with blockchain as its underlying infrastructure from the beginning, 6G and beyond networks can be more scalable and provide better and more efficient solutions in terms of spectrum sharing and resource management, for example.
Moreover, with privacy in mobile networks becoming more and more critical, due to the emergence of novel applications such as automated vehicles, industry 4.0 and medical applications, where even a minor failure can lead to disastrous consequences, blockchain can be of great advantage in securing and storing sensitive information.
Since all information in a blockchain is verified by all peers and it is immutable, this can allow future mobile networks to have a permanent record of all events with its corresponding time-frame~\cite{Kotobi2018}.
	
In contrast to other papers in the area, which analyze the impact of applying blockchain in wireless networks and spectrum management~\cite{Kotobi2018,Weiss2019,Dai2019}, in this article, we dive deeper in the field of blockchain-enabled resource sharing and spectrum management.
Based on that, in this paper a blockchain 6G-enabled resource management, spectrum sharing and computing and energy trading is envisioned as an enabler for future use-cases.
These resources are considered to be in a resource pool, in which spectrum is dynamically allocated, network slices are managed and hardware is virtualized in order to enable the blockchain resource and spectrum management.
Based on this envisioned framework, a discussion on how blockchain can enable resource sharing between devices, such as energy, data, spectrum lease, and computing power is presented.
In addition, the motivations of utilizing blockchain for different use-cases are highlighted, mainly in terms of the Internet of things (IoT) and D2D communications, network slicing, and network virtualization.
Lastly, some future trends expected in the realm of blockchain-enabled wireless networks are discussed and conclusions are drawn.
	
The remainder of this paper is organized as follows.	
Section~\ref{sec:Spectrum} presents an overview of current spectrum management and allocation techniques, as well as a link between blockchain and spectrum management.
Section~\ref{sec:Blockchain} discusses the motivations behind blockchain, and overviews its fundamentals.
Section~\ref{sec:App} discusses some key applications of blockchain and how it can transform current wireless networks.
Lastly, Section~\ref{sec:Conc} concludes the article.

\section{Spectrum Management}
\label{sec:Spectrum}
To meet the increasing high data rate demand of 5G and beyond applications, the capacity of the networks has to increase, hence, there is also an increase in the demand for spectrum. A dynamic policy for the management of the spectrum license has recently been proposed to manage the spectrum efficiently~\cite{Li2017}. It allows unlicensed secondary users to opportunistically access the licensed spectrum without interfering with the licensed primary user.
One of the options for using the new spectrum license is to distribute operation parameters to policy-based radio via a database. Such model has been established for sharing the Television White Space (TVWS) and the Citizen Broadband Radio Service~\cite{FCC2015}. Recently, the application of blockchain as trusted database has emerged~\cite{Liang2020} where information such as spectrum sensing and data mining outcomes, spectrum auction results, spectrum leasing mappings and the idle spectrum information are securely recorded on the blockchain. Blockhain thus brings new opportunities to Dynamic Spectrum Management (DSM)~\cite{Weiss2019,Dai2019,Liang2020} and it has recently been identified as a tool to reduce the administrative expenses associated with DSM~\cite{FCC2018}.  In particular, the features of blockchain can improve conventional spectrum management approaches such as spectrum auction~\cite{Kotobi2018}. Further, blockchain can aid in overcoming the security challenges and the lack of incentive associated with DSM~\cite{Liang2020}. Since blockchain is a distributed database, it lends this property such that records in the DSM system are recorded in a decentralized manner.

One of the key areas where blockchain finds application in spectrum management is in recording its information. Note that blockchain can record information as transactions, while spectrum management relies on databases such as the location-based database for protecting the primary users in the TVWS \cite{Gurney2008}. With blockchain, information about spectrum management such as 1) the TVWSs; 2) spectrum auction results; 3) the spectrum access history; and 4) the spectrum sensing outcomes, can be made available to the secondary user.
As such, the benefits of recording the spectrum management information with blockchain are discussed here
\begin{itemize}
    \item
    Contrary to conventional third party databases, blockchain enables users to have direct control of the data in the blockchain, thus guaranteeing the accuracy of the data. In particular, information on TVWS and other underutilized spectrum can be recorded in a blockchain. Such data could include the usage of the spectrum in frequency, time and the geo-location of TVWS, and the primary users' interference protection requirement.
    
    \item Improved efficiency of spectrum utilization with efficient management of the  secondary users' mobility, and the primary users' varying traffic demands. This is supported by the decentralized nature of the blockchain with primary users recording information on idle spectrum, which can be readily accessed by  unlicensed secondary users. Moreover, secondary users can make their arrival into the network or departure from it known to other users by initiating a transaction.
    
    \item Access fairness can be achieved with blockchain based approaches where access history is recorded. This is not the case with the traditional Carrier Sensing Multiple Access (CSMA) schemes where access is not coordinated. Access can be managed in blockchain via smart contracts, where a threshold is defined and users can be denied access to a specific band for a specified period of time when they reach the pre-defined access threshold.
    
    \item Blockchain provides a secure and verifiable approach to record information related to spectrum auction. Spectrum auction has been established as an efficient approach for dynamic allocation of spectrum resources \cite{Zhang2013}. The benefits of the blockchain based approach include 1) it prevents frauds from the primary users by providing transparency; 2) it guarantees that the auction payments are not rejected, since all transactions are verified before they are recorded on the blockchain; and 3) it prevents unauthorized secondary users from accessing the spectrum since all secondary users can cooperatively/collaboratively supervise and prevent such unauthorized access.
\end{itemize}
   
In \cite{Weiss2019}, the authors  explored the applications of blockchain in spectrum management including primary cooperative sharing, secondary cooperative sharing, secondary non-cooperative sharing and primary non-cooperative sharing. Moreover, in \cite{Kotobi2017}, the authors utilized a blockchain verification protocol for enabling and securing spectrum sharing in cognitive radio networks. Spectrum usage based on the blockchain verification protocol was shown to achieve significant gain over the traditional Aloha medium access protocol.  The authors in \cite{Qui2019} proposed a privacy preserving secure spectrum trading and sharing scheme, which was based on blockchain technology, for unmanned aerial vehicle  (UAV) assisted cellular networks. Furthermore, in \cite{Chai2019}, the authors proposed a consortium blockchain-based resource sharing framework for V2X which couples resource sharing and consensus process together by utilizing the reputation value of each vehicle.
In \cite{Dai2019}, the authors proposed the integration of blockchain technology and artificial intelligence (AI)  into wireless networks for flexible and secure resource sharing. 

\section{Benefit of Using Blockchain}
\label{sec:Blockchain}

\begin{figure*}[t!]
    \centering  
    \includegraphics[width=0.97\textwidth,trim={0cm 1cm 0cm 1cm},clip]{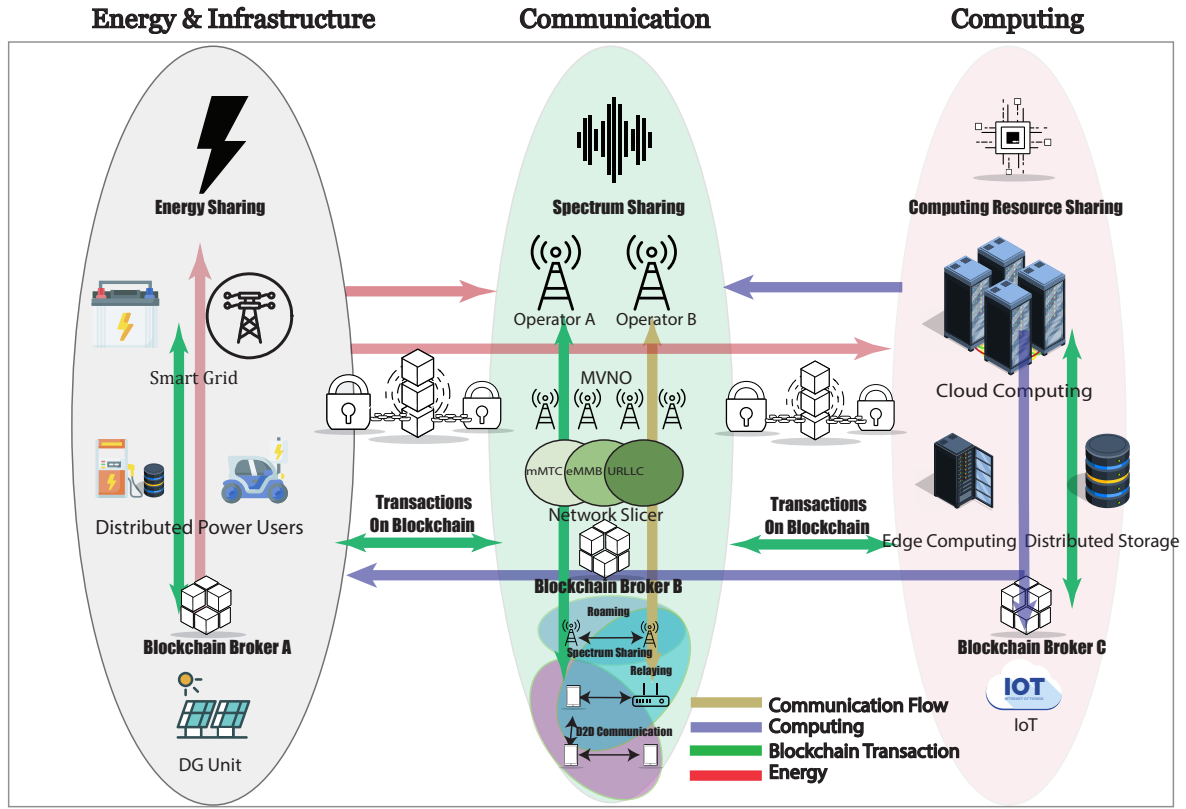} 
    \caption{Blockchain-enabled resource management framework}
    \label{fig:eco}
\end{figure*}

\subsection{Blockchain Basis}
Blockchain has a huge boost in the cryptocurrency and ledger keeping industry, and thanks to the vitality of the community, the technology has gained much attention from policymakers, mobile operators, and infrastructure commissioners~\cite{Liu2018}. 
Blockchains are distributed databases organized using a hash tree\footnote{A hash tree or Merkle tree is a tree in which every leaf node is labeled with the hash of a data block, and every non-leaf node is labeled with the cryptographic hash of the labels of its child nodes~\cite{Merkle1990}}, which is naturally tamper-proof and irreversible \cite{Crosby2016}.  
It has the attribute of adding distributed trust, and it is also built for enabling transaction consistency in a database. Furthermore, blockchain allows for atomicity, durability, auditability, and data integrity \cite{Bhattacharya2019}.  
Besides its chain-link data structure nature, the Consensus Mechanism (CM), which ensures a unambiguous ordering of transactions and guarantees the integrity and consistency of the blockchain across geographically distributed nodes, is of great importance to blockchains.
The CM largely determines blockchain system performance, such as transaction throughput, delay, node scalability, and security level, etc. As such, depending on application scenarios and performance requirements, different CMs can be considered.
Commonly used CMs include Practical Byzantine Fault Tolerance (PBFT), Proof of Work (PoW), or Proof of Stake (PoS), a detailed analysis of performance and security of consensuses and how they can be used in different resource management and sharing scenarios are stated in Section \ref{sec:cm}. 

\begin{table*}[ht!]
\centering
\caption{Comparison of blockchain consensus}
\begin{threeparttable}[h!]
\begin{tabular}{|c|c|c|c|c|c|c|c|}
\hline
Consensus & \begin{tabular}[c]{@{}c@{}}Suitable Type\\ of Blockchain\end{tabular} & Latency/TPS &BFT$^*$& \begin{tabular}[c]{@{}c@{}}Communication\\ Complexity$^\dag$\end{tabular} & \begin{tabular}[c]{@{}c@{}}Security\\ Threshold$^\ddag$\end{tabular} & \begin{tabular}[c]{@{}c@{}}Energy\\ Usage\end{tabular} & Scalability \\ \hline

PBFT & \begin{tabular}[c]{@{}c@{}}Consortium /\\ Private\end{tabular} & Low/ High\cite{Castro1999} & Yes \cite{Castro1999}& $O(n^2)$\cite{Castro1999} & 33\%\cite{Castro1999} & Low & Low \cite{Castro1999}\\ \hline

RAFT & \begin{tabular}[c]{@{}c@{}}Consortium /\\ Private\end{tabular} &\begin{tabular}[c]{@{}c@{}}Very Low /\\ Very High\cite{Xu2020}\end{tabular}  &No\cite{Xu2020}&  $O(n)$\cite{Ongaro2014} & 50\%\cite{Xu2020} & Low & Medium\cite{Ongaro2014} \\ \hline 

\begin{tabular}[c]{@{}c@{}}PoW /\\ PoS\end{tabular}  & Public & High/ Low \cite{nakamoto2019}&Yes\cite{niu2019}& $O(n)$ \cite{nakamoto2019}& 50\% \cite{nakamoto2019}& High& High\cite{nakamoto2019} \\ \hline

\begin{tabular}[c]{@{}c@{}}Proof of\\ Storage\end{tabular} & Public & High /Low \cite{Benet2014} & Yes\cite{niu2019} & $O(n)$ \cite{Benet2014}& 50\% \cite{Benet2014}& Low & High \cite{Benet2014}\\ \hline
\end{tabular}%
\begin{tablenotes}
\item $^*$The ability to tackle byzantine fault.
\item $^\dag$ $n$ indicates the number of participants.
\item $^\ddag$The given percentages stands for the maximum acceptable faulty nodes or attack.

\end{tablenotes}
\end{threeparttable}
\label{tab:cm}
\end{table*}
\vspace*{.2cm}

The blockchain opens up transparent and distributed information reformation, which can benefit all aspects of industries, accommodating all range of centralization using different CMs. 
In perspective of using blockchain technology in 6G, the massive deployment of blockchain may lead to a major step forward for the communications industry and all other departments of the economy. 

The transparent information flows on the blockchain are not only valuable assets for both users, and operators, but also to service providers and societies. In social practice, the authority has always attempted to grip every detail for every operation and transaction. However, it would never track down every happened transaction if they are not born to be recorded. 
Blockchain occurs to be the ideal tool for tracking of transactions if the blockchain native transactions are de facto in panoptic scenarios. The blockchain native resources and assets will stimulate a new era of information revolution. Such reformation will significantly improve the system efficiency and security thanks to the better public order~\cite{Foucault1977}.
It enables the Infrastructure as a Service (IaaS), Blockchain as a Service (BaaS) \cite{Singh2018} to spread out in terms of feasibility, and now the infrastructure can be organized in a distributed way by allowing the infrastructure to trade without further efforts to be centrally managed.

Later, such an ecosystem incubates the Blockchain as an Infrastructure (BaaI), which provides a solid tool-chain for settlements between the producer, the trader, and the consumer, as shown in Fig.~\ref{fig:eco}. 
As seen in Fig.~\ref{fig:eco}, blockchains can be the information backbone of a local distributed resource management system in the way of organizing the customers and producers in an open, transparent market, breaking up the information barriers to publicize the resources, and accelerate the flow of transactions.


Blockchain has incubated the new horizon of resource trading for fixed assets, such as licensed spectrum and computing hardware. In our proposed blockchain 6G resource management scheme, trade-able spectrum, and computing resources are integrated parts of resource pools, where spectrum is dynamically allocated, network slices are managed, and the hardware is virtualized in order to facility blockchain-enabled resource management.
The automated blockchain-enabled resource management relies on the programmable blockchain feature, which in most cases are described as smart contracts\footnote{The smart contract is essentially an executable program code stored on the chain, representing terms of agreements triggered automatically when certain conditions are met~\cite{Werbach2018}.}. The contract's content is transparent for both public and agreement making parties, which makes it publicly traceable. The virtual machine concept is used in the smart contract executions, where the code will be executed by a node on the virtual stack, and its results will be stored on the chain as a transaction record. The temper-proof ability and fully automated process give the contract high immutability against breaches of the contract and misrepresentations.

\subsection{Impact of Consensus and Security Performance}
\label{sec:cm}


If the impressive and resistive data structure of blockchain is the facade of a building, the consensus is the pillars. 
Blockchain has various options on the CM. Choosing a suitable consensus for 6G resource management is the most critical step of making a secured and efficient future-proof blockchain system. As the CM, which ensures an unambiguous ordering of transactions and guarantees the integrity and consistency of the blockchain across geographically distributed nodes, is of importance to blockchains since it determines its performance in terms of TPS, delay, node scalability, security, etc. 

Depending on the access criteria, the chain can be divided into public and private. 
The public chain is permission-less, which uses proof-based consensus to provide a secured, reliable network for every participant without requiring their identities at entry points. In the 6G resource pool, there are potential anonymous clients and providers on ad-hoc basis~\cite{Zhang2019}. 
The benefit of adopting a public chain is significant for ad-hoc networks, where the barriers of identification and security are broken down for panoptic information exchanges.
As such, public chains can potentially promotes the efficiency of the community and regulate the order of participants \cite{Foucault1977}.
However, if participants are concealed, violations and malicious activity are emerging threats to the system. 
The consortium/private chain, in contrast, is permissioned, meaning that entry is controlled.
It has a rather stable community composition, where the identity of participant is not kept secret.
The network faces fewer threats from unknown attacks but has challenges within the network, for instance, the malicious byzantine node\footnote{A byzantine node is a malicious node that conceals its existence and tempers the consensus, which tampers with the security of network. }.

Before adapting to any new technologies, security and reliability are always the principle concerns. Blockchain technology is born to outperform existing solutions regarding security performance and robustness. Table \ref{tab:cm} shows the comparison between widely used CMs of blockchain regarding 6 aspects: latency, TPS, complexity, security, energy consumption, and scalability. As it can be seen, private/consortium consensuses show better latency, TPS and energy consumption performance alongside lower ability to scale up, however, the applied application prioritize latency and TPS over scalability. On the other hand, proof-based mechanisms have decent performance on scalability, but sacrifice latency and TPS.
In some cases like proof of work, it also consumes a huge amount of power. 
However, their good scalability gives them capability to grow fast in the public network and does not suffer from a surge of users, which makes them excellent at mass market trading and distributed file storage system. Regarding the security performance, it is worth noting that the non-byzantine consensuses assume non-malicious activities, but byzantine consensus has tolerance not only against inactivity, but also against false and erroneous messages. PBFT functions with less than $(n-1)/3$ byzantine nodes, and some variants of PBFT provides higher tolerance with trades-off of latency, such as multi-layer PBFT.


Besides the consensus which secures the blockchain from top-level threats, the communication link should be hardened to prevent external security breaches. The wireless communication is in peril of jamming and spoofing because of open channels. In the practice of wireless blockchain network, the communication failure will result the node failure and thus lower the security level. To mitigate the transmission success rate, collision avoidance mechanism like Carrier Sense Multiple Access with Collision Avoidance (CSMA/CA) and physical layer security can be considered.


\section{Application Scenarios}
\label{sec:App}
\subsection{IoT and D2D Communications}
The IoT is a paradigm which envisions that all our daily objects and appliances will be connected to each other, collecting and sharing information.
This will allow the automation of certain tasks and enable several other applications to emerge, such as smart homes, smart transportation and wearable devices, smart farming, healthcare, machine-to-machine communications, etc.~\cite{Carames2018}.
In order to reach such automation and growth, it is necessary to have proper standards and protocols for IoT devices.
However, current solutions still rely on a centralized model, which incurs in a high maintenance cost for manufacturers, while consumers also lack trust in these devices.
That, combined with the resource constraints of IoT devices, privacy and security concerns, as well as poor interoperability among different vendors, makes IoT a challenging domain~\cite{Christidis2016,Cao2019}.
Similarly, D2D communications, a paradigm that envisions the communications and share of data between devices, also shares similar challenges to the IoT~\cite{Fodor2012}.
For example, mobile devices are constrained by battery, while security is an ever-present concern in mobile communications.
Moreover, in order to fully enable D2D communications, a proper incentive for trading and sharing resources, such as power or data is needed, as current D2D paradigms lack motivation to do so~\cite{Fodor2012}.
	
In this context, blockchain is an excellent compliment to both IoT and D2D communications, as it can provide the underlying infrastructure with improved interoperability, privacy, reliability and scalability~\cite{Cao2019}.
In the context of resource management, for example, blockchains can be used to perform spectrum sharing and record all the spectrum utilization and lease requests~\cite{Weiss2019}.
Moreover, it can provide the incentive needed for devices to share and trade resources, as current protocols lack the incentive to do so.
By integrating blockchain into IoT and D2D, it can provide rewards every time devices share their power or data, allowing for a more cooperative and trusted network environment~\cite{Cao2019,Liu2018}.
Moreover, this reward mechanism can also be applied in the context of spectrum sharing, in which whenever a user leases spectrum to another, a reward can be assigned, creating a more collaborative environment and improving spectrum efficiency~\cite{Kotobi2018,Weiss2019}.
Furthermore, blockchains can be utilized in the realm of Vehicular-to-Anything (V2X) communications, by encouraging vehicles to trade energy or information with each other~\cite{Dai2019}.
In addition, another key aspect of V2X communications is how to guarantee a secure communication between vehicles and public key infrastructures (PKI).
In this context, blockchain can be utilized as the infrastructure to provide secure and private communications to the PKI, or also the communication between PKIs from different vendors~\cite{Liu2018}.
	
However, despite all of these benefits, the integration of blockchains in the IoT and D2D domain is still challenging~\cite{Dai2019Survey,Dai2019,Weiss2019}.
In the case of public chains, for example, the decentralized CMs often require extensive computing power from network nodes (such as PoW based blockchains).
This can be a problem, as most IoT devices are power constrained.
This is specially true for devices powered by cellular IoT, which can be deployed in very remote or difficult to access areas and are expected to have over ten years of battery life~\cite{mangalvedhe2016nb}. 
Thus, the utilization of blockchain in cellular IoT, especially when considering the computation of the consensus algorithm, can significantly reduce the life-time of cellular IoT devices, limiting their communication capabilities and effectiveness.
As such, it is still not clear how the generation of the PoW could be done when integrating public blockchains with IoT or D2D communications~\cite{Cao2019}.
Hence, other CMs such as PBFT are being proposed in the context of IoT applications~\cite{Cao2019,Onireti2019}.
Another challenge in integrating blockchain to small devices comes due to their limit memory capabilities.
Since in blockchain every node needs to have a record of all the current and previous blocks in the chain, it can be infeasible to store such a huge amount of data in IoT devices. 
Thus, it is still not clear how blockchain can be fully integrated in IoT.
Moreover, blockchain still has privacy issues, as it has been shown by other studies that by analyzing transaction patterns, identities of users could be inferred~\cite{Dai2019Survey}.

On top of that, it is also known that blockchain introduces delay, due to its decentralized approach and its CMs.
As such, this additional delay might also impact the performance of certain wireless communication use-cases, such as in V2X, industrial applications, or D2D and it is still an area to be investigated.
Moreover, in V2X scenarios, information security and resilience are critical, since any small failure can lead to catastrophic and even fatal consequences.
As such, in those cases, blockchain can provide an additional secure layer in order for vehicles to perform key management exchange, as in~\cite{lei2017blockchain}, or even to protect a vehicle's identity and location, in what is known as pseudonym management~\cite{bao2019pseudonym}.
Lastly, another important challenge in this realm, which has not been largely explored is how the performance of the wireless link affects the performance of blockchain~\cite{Sun2019b}.
Despite recent works investigating the suitability of the CSMA/CA protocol in wireless blockchain networks~\cite{Cao2020}, or the security performance and optimal node deployment of blockchain-enabled IoT systems~\cite{Sun2019}, more research needs to be done in this area.

\subsection{Network Slicing}
Network slicing is a very promising technology in future cellular architecture, and it is aimed at meeting the diversified requirements of different vertical industry services. Network slicing is a specific form of virtualization that allows multiple logical networks to run on top of a shared physical network infrastructure \cite{Foukas2017}. A network slice is realized when a number of Virtualized Network Functions (VNF) are chained based on well defined service requirements, such as the massive Machine Type Communication (mMTC), enhanced Mobile Broadband (eMBB) and the ultra-Reliable Low Latency Communication (uRLLC). 
 The management and orchestration of network slices must be trusted and well secured, in particular for accommodating applications that require high security, such as in the case of remote robotic surgery and V2X communications 
 \cite{Chang2019}.
 
\begin{figure}[t!]
    \centering
    \includegraphics[width=0.95\columnwidth,trim={1.3cm 1.3cm 1.2cm 1.3cm},clip]{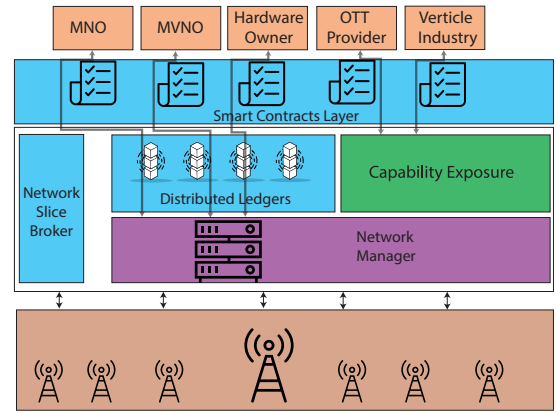}
    \caption{Spectrum management using blockchain and smart contract.}
    \label{fig:BCandSpectrum}
\end{figure}

Network slicing also enables Mobile Network Operators (MNO) to slice a single physical network into multiple virtual networks which are optimized based on specified business and service goals \cite{Alliance2016}. Hence the term Mobile Virtual Network Operators (MVNOs). The implementation of MVNOs necessitate the integration of a network slice broker into the architecture, as seen in Fig.~\ref{fig:BCandSpectrum}. 
\subsubsection{Network Slicing Broker}
The aim of a network slice broker is to enable MVNOs, industry vertical market players, and  Over-The-Top (OTT) providers to dynamically request and release the network resources from the infrastructure provider entity based on their needs \cite{Samdanis2016}. The network slicing brokering relies on the ability of the MNO/Communication Service Providers (CSP) to automatically and easily negotiate with the requests of the external tenants of the network slice based on the currently available resources with the infrastructure provider. In \cite{Samdanis2016}, the authors proposed the concept of 5G network slice broker that could lease network resources on-demand basis.

The 3GPP's study on orchestration and management of network slicing for 5G \& beyond networks indicated the establishment of mutual trust between the actors (MVNOs, MNOs, OTT providers) as a prerequisite for an effective and efficient multi-operator slice creation \cite{3GPPTR28801}. Hence, trust and security are important factors to be considered in the implementation, design and integration of a network work slice broker.

\subsubsection{Integration of Blockchain to Network Slicing and Resource Brokerage}
A major challenge associated with network slicing and resource brokerage is the need to keep a transparent, fair and open system within the available number of resources and several suspicious players. 

Blockchain and Distributed Ledger Technology (DLT) functionalities can be utilized to address the aforementioned trust and security issues associated with the implementation of network slicing either for the coexistence of various application and services, or for both the service and operational use-cases of CSPs.  The trading of network slice can be blockchain-based where the blockchain smart contract orders the slice orchestration based on the agreed SLA from the 5G network slice broker.  Blockchain can be integrated for taking the record of how each resource has been used and how each service provider has performed against the SLA. Blockchain combines a distributed network structure, CM and advanced cryptography to present promising features which are not available in the existing structures. The key gain that is achieved through blockchain is the integration of the trust layer which lowers the collaboration/cooperation barrier and enables an effective and efficient ecosystem. Further the distributed nature of blockchain prevent the single point of failure problem and thus enhance security. 

\begin{figure}[t!]
\centering
    \includegraphics[width=0.97\columnwidth,trim={0.3cm 1.8cm 0.3cm 1cm},clip]{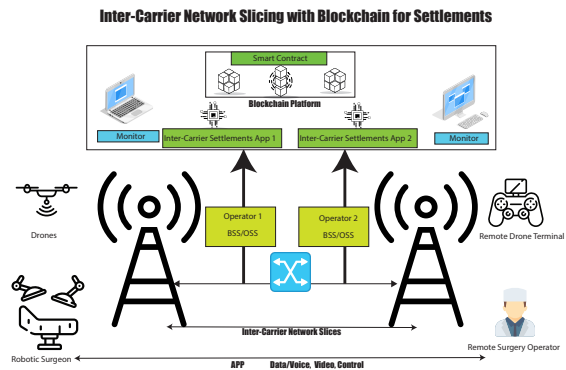}
    \caption{Network slicing applied with use of blockchain.}
    \label{fig:DraftNetworkSlicer2}
\end{figure}

Fig.~\ref{fig:DraftNetworkSlicer2} illustrates the provision of remote surgery/consultation and remote control of drones over a long distance (with network operators in different geographies), while leveraging on network slicing and blockchain technologies. Here, a blockchain based approach is used to automate the reconciliation and the payment between provider in different geographies. Without this approach a more costly manual intervention or the integration of a third party for settlement would be required.
Blockchain can also enable seamless access of devices to a diverse number of networks. However, this might require the network provider to manage rules, agreement and transactions across a rising number of access points. Blockchain can play a reinforcing role such as in the case of auditing agreement. Once information is stored on a blockchain, it can be acted upon through ``smart contracts'' \cite{Crosby2016}.

In \cite{Backman2017}, the authors proposed a model where brokering is managed by the 5G network slice broker \cite{Samdanis2016} while the payout, billing and leasing are managed by the blockchain-based slice leasing ledger which is incorporated in the service layer.  
Blockchain can enable secure and automated brokerage of network slicing while proving the following gains:
\begin{itemize}
    \item Significant savings in the operational (transaction and coordination) cost;
    \item Speed up the slice negotiation process leading to a fall in the cost of slicing agreement;
    \item Increased efficiency of operation for each network slice \cite{Accenture2019};
    \item Increased security of the network slice transactions;
    \item The creation a blockchain-enabled contract for MVNOs and MNO that cannot afford the required network capital investment which could be on the high side. In particular, the frequency spectrum could be leased by large operators or players on a pay-as-you-go basis or in real time.
\end{itemize}
Blockchain can also enhance the enforcement of quite straightforward agreement which are related to many brokering operations. Furthermore, the negotiation on SLAs can be more efficient when pricing and Quality of Service (QoS) levels are identified as smart contract parameters.

Other opportunities associated with blockchain in the next generation networks include
\begin{itemize}
    \item The settlement of transaction between multiple carriers including voice transactions and Call Detail Records (CDRs) of all involved call participants;
    \item Managing the Service Level Agreement (SLA);
    \item Simplification of roaming terms and agreement between multiple operators;
    \item Managing money transfer across boarders and cross-carrier payment platform;
    \item Managing user/nodes identity and authentication process;
    \item Managing Licensed Spectrum Access (LSA) via the blockchain-based carrier marketplace.
\end{itemize}  

\subsection{Inter-domain Blockchain Ecosystem}

Shareable resources are the new assets defined by the distributed resources operators, which are not limited to communication but energy and computing sections. While communication infrastructure also relies on the energy and computing resource provision, as shown in Fig.~\ref{fig:eco}. Thus, a trusted blockchain-enabled trading ecosystem including energy, computing and communication can be built to enable an efficient and sustainable 6G. 

In the ecosystem, we can find various streams of blockchain transaction, energy and computing flows using shared communication assets in the resource management scheme, as seen in Fig.~\ref{fig:eco}. Arrows in Fig.~\ref{fig:eco} represents the flow directions and they are started with the provider, through the inter-domain sharing scheme to reach the final consumers on both local level with consortium blockchain and national or global level via public blockchain. The ecosystem is not limited to the scope of energy, communication and computing as it can expand itself to wider range through cross-field integration to reach, for instance, automotive, finance, manufacturing, logistic chain and so on. 

Organizations who intend to fuse such resources can be recognized as Virtual Infrastructure Operators (VIO), since they do not own all of the resources but a vendor of combined sets of resources.
An example of VIO can be found in remote regions, where local infrastructure investors tend to have off-grid Distributed Generation (DG) units \cite{Talapur2018}, for instance, solar and wind farms and micro Combined Heat and Power (microCHP) to offer energy and heat to remote users in the form of Distributed Energy Resources (DER)~\cite{Landsbergen2007}. A local based integration of such resources, aka, Virtual Power Plant (VPP), plays the role of the vendors for electricity and heat and also buys from or sells to other grids, with unfilled demands and excess electricity.
Since these establishments are far away from the central network and lack of a cost-effective way of trading regarding the communication and delivery cost, it is ideal to broke with other local providers and exchange the electricity for other goods.
For example, the communication relay service and computing service for DG sensors in exchange of powering up the hardware, hence, the ecosystem is fostered while the internal demands grows.
Besides the operator owned resources, there are many common resource containers among all participants. 


However, blockchain ecosystem has to accommodate the performance and security requirement of the intended application. In terms of the performance and security, the consensus is the major concern in the phase of planning.
Different consensuses can be applied to the sharing scheme. For example, a public chain is more suitable to inter-domain transactions on top level operators like the national grid and first tier MNO. However, if the resources are local-oriented, the private chain can be hosted for IoT and local/off-grid nodes, where the information from a private chain is kept within the network with confidence for external auditing. An ecosystem may introduce multiple consensuses on different chains to achieve its best results.

Beyond the deployment of blockchain, the actual hardware plays important roles in the ecosystem, as current blockchain applications are designed for upper layer applications, it lacks of understanding of the portable solutions for mobile device, such as drones, cars and IoT. It is worth noting that the wireless capability for blockchain is essential in 6G deployment. Wireless blockchain-enabled nodes empower the Machine-to-Machine (M2M) trade among distributed and shared resources, therefore it becomes essential that the remote nodes are wireless enabled. In the near future, the VANET-enabled car equipped with blockchain nodes can recharge the battery from multiple wireless charging points while moving and trade the information it carries, for instance, the Light Detection and Ranging (LiDAR) mapping data, relaying the internet access, edge computing resource and anything that can be used by the remote DG unit using wireless communication, D2D, and edge computing. The transactions are kept in the blockchain and carried by the vehicular network then mined by the local infrastructure or base station blockchain nodes. Later, the mined blocks will be relayed by satellite-linked base stations for a fee~\cite{Yaacoub2019}. The auction of spectrum and network slices can be found on data relay and short-range Vehicle to Ground (V2G) communication, which requires huge local bandwidth to achieve lower latency. This example intends to give an insight of inter-domain blockchain ecosystem, and further additional features are all made possible based on the inter-domain transactions.

\subsection{Challenges of Applying Blockchain Technology in Resource Sharing and Spectrum Management}
Though blockchain has many advantages, some features need to be eliminated when applied to the resource sharing and spectrum management scenarios. Here we highlight some of the challenges of applying blockchain technology in resource sharing and spectrum management.

 Storage:  Each replica node in the conventional blockchain network must process and store a copy of the completed transaction data. This can give rise to both storage and computation burden on IoT devices, which are generally resource constrained, thus limiting their participation in the blockchain network. 

Underlying Networking: Implementing a consensus mechanism within the blockchain is computationally expensive and it also requires significant bandwidth resources. Meanwhile, resources are very limited in future network thus meeting the resource requirement for large transaction throughput might be hard to achieve with the current system. 

 Scalability of the Blockchain Network: The scalability of the blockchain network is a serious issue in current systems. The number of replicas in the blockchain network relates directly to the throughput (i.e., number transactions per second) and latency (i.e., time required to add a transaction to the blockchain). Hence, sustaining the huge volume of transactions expected in blockchain-enabled future networks demands solutions on improving the throughput of the blockchain system.

\section{Conclusion}
\label{sec:Conc}
In this article, a blockchain-enabled 6G resource management, spectrum sharing, and computing and energy trading was envisioned as an enabler for future use-cases.  We first briefly introduced the current spectrum management and allocation techniques and discussed the link between blockchain and spectrum management. We have then given the motivation behind blockchain as well as an overview of its fundamentals. Moreover, we have discussed some of the key application of blockchain and the transformation that it brings to the current wireless networks. The discussed applications include IoT and D2D communications, network slicing, and the inter-domain blockchain ecosystem.

To enable the full ecosystem and manage the resource for 6G, we have identified the following open problems: 1) development of lightweight blockchain solutions for low-cost IoT devices; 2) high-performance blockchain and decentralization for the vertical industries and future networks; 3) Developing blockchain solutions ecosystem by considering the security and privacy issues; 4) implementation of blockchain protocols over the wireless channel and evaluating fundamental limits relating to the performance and security. 

\bibliographystyle{elsarticle-num}
\bibliography{egbib}

\end{document}